\documentclass[aps,pra,superscriptaddress,showpacs,twocolumn]{revtex4-1}
\usepackage{graphicx}
\usepackage{epsfig,amsmath}
\usepackage{subfigure}
\usepackage[colorlinks=true,linkcolor=blue,urlcolor=black,citecolor=blue]{hyperref}
\usepackage{dcolumn}
\usepackage{stmaryrd}
\usepackage{mathrsfs}
\usepackage{pifont}
\usepackage{amsthm}
\usepackage{amssymb}
\usepackage{bm}
\usepackage{amsmath}
\usepackage{pdfpages}
\newcommand{\Qnm}{Q_m^{(n)}}
\newcommand{\Qn}{{Q_0^{(n)}}}
\newcommand{\sigmazav}{\langle  \sigma_z \rangle}
\newcommand{\NQmax}{\mathcal{N}}
\newcommand{\av}[1]{\langle #1 \rangle}
\newcommand{\barg}{\av{g_k}}

\begin{document}

\title{Approach to solving spin-boson dynamics via non-Markovian quantum trajectories}

\author{Zeng-Zhao Li}
\thanks{Z.Z.L and C.T.Y contributed equally to this work.}
\affiliation{Beijing Computational Science Research Center, Beijing 100084,
China}
\affiliation{Department of Applied Physics, Hong Kong Polytechnic University, Hung Hom,
Hong Kong, China}
\author{Cho-Tung Yip}
\thanks{Z.Z.L and C.T.Y contributed equally to this work.}
\affiliation{Department of Applied Physics, Hong Kong Polytechnic University, Hung Hom,
Hong Kong, China}
\author{Hai-Yao Deng}
\affiliation{Department of Applied Physics, Hong Kong Polytechnic University, Hung Hom,
Hong Kong, China}
\affiliation{International Center for Materials Nanoarchitechtonics, National Institute for Materials Science, Namiki 1-1, Tsukuba 305-0044, Japan}
\author{Mi Chen}
\affiliation{Beijing Computational Science Research Center, Beijing 100084, China}
\affiliation{Department of Physics, Fudan University, Shanghai 200433, China}
\author{Ting Yu}
\affiliation{Beijing Computational Science Research Center, Beijing 100084,
China}
\affiliation{Center for Controlled Quantum Systems and Department of Physics and Engineering Physics, Stevens Institute of
Technology, Hoboken, New Jersey 07030, USA}
\author{J. Q. You}
\thanks{Corresponding authors: J. Q. You (jqyou@csrc.ac.cn) and C. H. Lam(C.H.Lam@polyu.edu.hk)}
\affiliation{Beijing Computational Science Research Center, Beijing 100084,
China}
\affiliation{Synergetic Innovation Center of Quantum Information and Quantum Physics, University of Science and Technology of China, Hefei, Anhui 230026, China}
\author{Chi-Hang Lam}
\thanks{Corresponding authors: J. Q. You (jqyou@csrc.ac.cn) and C. H. Lam(C.H.Lam@polyu.edu.hk)}
\affiliation{Department of Applied Physics, Hong Kong Polytechnic University, Hung Hom,
Hong Kong, China}
\date{\today}

\begin{abstract}
We develop a systematic and efficient approach for numerically solving the non-Markovian quantum state diffusion
equation for an open quantum system that can be strongly coupled to an environment.
As an important application, we consider a real-time simulation of a spin-boson model in a strong coupling regime that is difficult to deal with using conventional methods. 
We show that the non-Markovian stochastic Schr\"{o}dinger equation can be efficiently implemented as a real--time simulation for this model,
so as to give an accurate description of spin-boson dynamics beyond
the rotating-wave approximation.
\end{abstract}

\maketitle
\section{Introduction \label{sec:intro}}
Dynamics of open quantum systems has been extensively studied in the last decades due to its pivotal importance in the areas of quantum optics, quantum dissipative dynamics and quantum information \cite{ScullyZubairy1997,Legget1987RMP,NielsenChuang2000}. 
The Lindblad master equations under the Born-Markov approximations are the major theoretical tools in depicting quantum evolution under the influence of external noises, but they are doomed to fail when the system-environment coupling becomes strong or when the environment is a structured medium \cite{bandgap}. 
Moreover,
the widely used rotating-wave approximation (RWA) 
ceases to be valid at a strong coupling regime
\cite{ScullyZubairy1997,EmaryBrandes2003PRL,Liberato2007PRL,LiParaoanu2013Ncommun}. It becomes clear that to correctly explain the novel quantum-mechanical phenomena arising from the strong-coupling physics,
the counter-rotating terms neglected in the RWA must be taken into account properly.  In addition, the counter-rotating terms
are known to be important in understanding quantum Zeno and anti-Zeno effects \cite{KofmanKurizki2004PRL,ZhengZhuZubairy2008PRL,CaoYouNori2010PRA}.
All the current researches going beyond the RWA and Markov approximation have shown the necessity of developing a powerful approach to dealing with new physics
arising from the strong coupling between the open quantum system of interest and its environment \cite{ZhengZhuZubairy2008PRL,CaoYouNori2010PRA,Agarwal2012PRA,Sornborger2004PRA,Werlang2008PRA}.


A stochastic Schr\"{o}dinger equation named the non-Markovian quantum state diffusion (QSD) equation
derived from a microscopic model has several advantages over the exact master equations.
While the exact master equations exist only for a few solvable models
(see, e.g., Ref.~[\onlinecite{Hu1992PRD}]), the exact QSD equation has been established for a generic class
 of quantum open systems  \cite{DiosiGisinStrunz1998PRA}. However, the applications of
 the exact QSD equation are severely limited unless this time-nolocal integro-differential equation
 can be cast into a numerically implementable time-local form
  \cite{DiosiGisinStrunz1998PRA,Yu1999PRA,StrunzYu2004PRA,JingYu2010PRL}. 
In the real-world problems, solving the exact dynamical equations in a strong coupling regime is very difficult. Therefore, it is imperative to develop an efficient perturbative
method that can be implemented to solve open system dynamics dictated by the strong coupling and structured medium.


In this 
paper, we develop a systematic and efficient approach to solving the non-Markovian QSD equations for open quantum systems up to arbitrary orders of noises. 
The major breakthrough is to convert the non-Markovian QSD equation into a set of coupled stochastic ordinary differential equations (SODE's) which efficiently evaluates a series expansion of the previously unsolvable $O$-operator up to arbitrarily high orders. The method can be generally applied to an arbitrary finite-state open system coupled to a bosonic bath with a Lorentzian noise spectrum at zero temperature.  As an important example, our method is used to solve a spin-boson model with a Lorentzian environment at zero temperature in the strong coupling regime that is previously intractable when real-time quantum dynamics is needed.

\section{Exact QSD equation \label{sec:intro}}
To put our discussion into perspective, we first consider a generic open quantum system with the following Hamiltonian (setting $\hbar=1$)\cite{DiosiGisinStrunz1998PRA}:
\begin{equation}
H_{\rm tot}=H_{\mathrm{sys}}+\sum_k (g_k L b_k^{\dagger}+g_k^{\ast}L^{\dagger}b_k)+\sum_k  \omega_k b_k^{\dagger}b_k,
\end{equation}
where $H_{\mathrm{sys}}$ is the Hamiltonian of the system under consideration, $L$ is the Lindblad operator, and $b_k$ denotes the annihilation operator of the $k$th mode of the bosonic bath.
The state of the bath may be specified by a set of complex numbers $\{ z_k \}$ labeling the (Bargmann) coherent states of all modes. The function $z_t$ that characterizes time-dependent states of the bath may be defined
by the Fourier expansion $z_t^{\ast}\equiv -i\sum_k g_k^{\ast}z_k^{\ast}e^{i\omega_k t}$.
When $z_k$ is interpreted as a Gaussian random variable, then $z_t$ becomes a Gaussian process with the correlation function obtained by the statistical mean
 $\alpha(t-s)=\av{z_{t}z_{s}^{\ast }} = \sum_k |g_k|^2 e^{-i\omega_k \left(t-s\right)}$.
For the simple case with a zero-temperature bath, the system state at time $t$ obtained from projecting the total state to
 the bath state  $|z\rangle$,  $\psi_t(z^\ast) \equiv  \langle z|\Psi_{\rm tot}(t)\rangle$, which is called a quantum trajectory, obeys a linear QSD equation \cite{DiosiGisinStrunz1998PRA} 
\begin{equation}
\dot{\psi}_t =-iH_{\mathrm{sys}}\psi_t +Lz_t^{\ast}\psi_t-L^{\dagger}\bar{O}
\psi_t.
\label{eq:linearSSE}
\end{equation}
Here, the $O$-operator is defined by ${\delta \psi_t}/{\delta z_s^{\ast}}=O\left( t,s,z^{\ast}\right)\psi_t$, and  $\bar{O}\left( t,z^{\ast}\right)=\int^{t}_{0}\alpha\left(t-s\right)O\left( t,s,z^{\ast}\right)ds$. 
Evaluating the $O$ operator poses a major challenge in solving quantum open systems in real-world applications. It is remarkable that the evolution is completely decoupled from projections to other bath states and hence can be solved independently.
In practice, one may adopt an importance sampling scheme in which the normalized system state $\tilde\psi_t(\tilde z^\ast) = \psi_t(\tilde z^\ast)/ |\psi_t(\tilde z^\ast)|$ is governed by the norm-conserving nonlinear QSD equation \cite{DiosiGisinStrunz1998PRA}, 
\begin{eqnarray}
\dot{\tilde{\psi}}_t&=&-iH_{\mathrm{sys}}\tilde{\psi}_t+\left(L-\langle L\rangle_t\right)\tilde{z}_t^{\ast}\tilde{\psi}_t \notag \\
&&-\left[\left(L^{\dagger}-\langle L^{\dagger}\rangle_t\right)\bar{O}-\langle\left(L^{\dagger}-\langle L^{\dagger}\rangle_t\right)\bar{O}\rangle_t\right]\tilde{\psi}_t,
\label{eq:nonlinearSSE}
\end{eqnarray}
where $\bar{O}$ denotes $\bar{O}\left(t,\tilde z^{\ast}\right)$ and $\langle ...  \rangle_t=\langle \tilde{\psi}_t|...|\tilde{\psi}_t\rangle$.
We define a shifted noise as
$\tilde{z}_{t}^{\ast }=z_{t}^{\ast }+ y_t$,  where the shift $y_t = \int_{0}^{t}\alpha ^{\ast}\left( t-s\right) \left\langle L^{\dagger }\right\rangle _{s}ds$ satisfies $y_0=0$,  and
\begin{equation}
\dot{y}_t=-\gamma y_t+\alpha^{\ast}\left(0\right)\langle L^{\dagger}\rangle_t.
\label{eq:shiftednoisediff}
\end{equation}
The state of the open quantum system at $t$, represented by the reduced density matrix  $\rho_t=\rm{Tr}_{\rm env} |\Psi_{\rm tot}\rangle \langle \Psi_{\rm tot}|$,
can be recovered from an ensemble average $\rho_t = \av{ |\tilde{\psi}_t\left(\tilde z^{\ast }\right)\rangle \langle \tilde{\psi}_t\left(\tilde z^{\ast }\right)|}$.


The QSD equations (\ref{eq:linearSSE}) and  (\ref{eq:nonlinearSSE}) are exact.  A key challenge is the determination of the $O$ operator contained in these equations.
For most practical problems except for a few specific examples where the exact $O$ may be explicitly determined \cite{DiosiGisinStrunz1998PRA,Yu1999PRA,StrunzYu2004PRA,JingYu2010PRL}, 
one has to resort to a  functional expansion \cite{Yu1999PRA} in terms of $\tilde z_{t}^{\ast}$ which after adapting to $\bar O$ writes
\begin{eqnarray}
\bar{O}\left( t,\tilde{z}^{\ast}\right) &=&\bar{O}^{\left( 0\right) }\left( t\right)
+\int_{0}^{t}\bar{O}^{\left( 1\right) }\left( t,\upsilon _{1}\right)
\tilde{z}^{\ast}_{\upsilon _{1}}d\upsilon _{1} \notag \\
&&+\int_{0}^{t}\int_{0}^{t}\bar{O}^{\left(
2\right) }\left( t,\upsilon _{1},\upsilon _{2}\right) \tilde{z}^{\ast}_{\upsilon
_{1}}\tilde{z}^{\ast}_{\upsilon _{2}}d\upsilon _{1}d\upsilon _{2}+\cdots \notag \\
&&+\int_{0}^{t}\cdots \int_{0}^{t}\bar{O}^{\left( n\right) }\left(
t,\upsilon _{1},\cdots ,\upsilon _{n}\right) \tilde{z}^{\ast}_{\upsilon _{1}}\cdots
\tilde{z}^{\ast}_{\upsilon _{n}} \notag \\
&&\times d\upsilon _{1}\cdots d\upsilon _{n}+\cdots ,  \label{eq:functional}
\end{eqnarray}%
where
$\bar{O}^{\left( n\right) }$
is symmetric with respect to the time variables $\upsilon _{i}$. However, finding $\bar{O}^{(n)}$ and performing the integrations for higher order terms
are formidable tasks and have only been performed up to $n\le 2$ for some specific models \cite{Jing2013}.



\section{SODE formulation}
%
In this work, we show that the QSD perturbation may be carried out to an arbitrary order of noise terms. Specifically, we can efficiently evaluate
Eq.~(\ref{eq:functional}) up to $\NQmax=100$ perturbative terms for the spin-boson model under consideration. We first rewrite it as
\begin{equation}
\bar{O}\left( t,\tilde{z}^{\ast}\right) =\sum_{n=0}^{N_{Q}}Q_{0}^{(n)}\left( t,\tilde{z}^{\ast}\right), \label{eq:barO}
\end{equation}
%
where $N_Q =\NQmax$ nominally but we allow $N_Q < \NQmax$ when higher order terms are vanishingly small.
We also define a generalized operator,
\begin{eqnarray}
Q_{m}^{(n)}\left( t,\tilde{z}^{\ast}\right) &=&\int_{0}^{t}\cdots \int_{0}^{t}\alpha \left(
t-\upsilon _{1}\right) \cdots \alpha \left( t-\upsilon _{m}\right)
\tilde{z}^{\ast}_{\upsilon _{m+1}}  \notag \\
&&\cdots \tilde{z}^{\ast}_{\upsilon _{n}}\bar{O}^{\left( n\right) }\left( t,\upsilon
_{1},\cdots ,\upsilon _{n}\right) d\upsilon _{1}\cdots d\upsilon _{n}.
\label{eq:Q_n_m}
\end{eqnarray}%
For $m\neq 0$, $\Qnm$ does not contribute directly to $\bar{O}$ but is an auxiliary operator needed to be solved simultaneously.
%
Let $\barg$ be a mean coupling strength.
Up to leading orders $\alpha(t) \sim \barg^2$, we have, $\tilde{z}^{\ast}_t \sim \barg$,  and hence
$Q_m^{\left(n\right)}\sim
\barg^{n+m+2}$ when using also $\bar{O}^{\left(n\right)} \sim \alpha(t)$ \cite{Yu1999PRA}.

%
From Eq.~(\ref{eq:Q_n_m}), each $\Qnm$ is a $n$-dimensional definite time-integral from 0 to $t$ in every dimension.  At time $t=0$, $\Qnm$ is exactly zero. For sufficiently small $t$, $\Qnm$ roughly scales as  $t^n$ assuming that the integrand varies smoothly with $t$. Then, $\Qnm \sim t^n \rightarrow 0$ for large $n$ and small $t$.
Therefore, the infinite series in Eq.~(\ref{eq:functional}) which involves only the $Q^{(n)}_0$'s in particular is then guaranteed to be convergent at least for small $t$.
More generally, $\Qnm$ has a finite support (i.e., a domain where  $\Qnm$ takes non-zero values) on the $(n, m)$-plane which expands with $t$.   Therefore, Eq.~(\ref{eq:functional}) and equivalently Eq.~(\ref{eq:barO}) can be arbitrarily accurate at a finite $N_Q$.
As $t$ increases especially for a strong coupling regime, the support might expand unboundedly. In practice, we impose the constraint $N_Q \le \NQmax$ by choosing a large $\NQmax$ to assure the accuracy, and consider $\Qnm$ only up to $n+m\le \NQmax$  corresponding to order $\av{g_k}^{\NQmax+2}$.  

%

For simplicity, we consider the environmental noise $z_{t}$ characterized by the Ornstein-Uhlenbeck
noise with the auto-correlation,
\begin{equation}
\alpha \left( t-s\right) =\frac{\Gamma \gamma }{2}e^{-\gamma \left\vert
t-s\right\vert }.
\label{eq:alpha}
\end{equation}
Taking the time derivative of Eq.~(\ref{eq:Q_n_m}) and applying the evolution equation of $\bar{O}^{(n)}$ \cite{Yu1999PRA},  we arrive at our central analytical result
after some algebra (see Appendix~\ref{sec:AppendA}) 
\begin{widetext}
\begin{eqnarray}
\dot{Q}_{m}^{(n)}
&=&\delta _{n,0}\alpha \left( 0\right) L+\frac{m}{n^{\prime }}\alpha \left(
0\right) \left[ L,Q_{m-1}^{(n-1)} \right] +%
\frac{n-m}{n^{\prime }}\tilde{z}_{t}^{\ast }\left[ L,Q_{m}^{(n-1)} \right] -\left( m+1\right) \gamma Q_{m}^{(n)}
-{i}\left[H_{\mathrm{s}},Q_{m}^{(n)} %
\right] \notag \\
&&-\sum_{k=0}^{n}\sum_{l=l_{a}}^{l_{b}}\frac{C_{l}^{k}C_{n-m-l}^{n-k}}{%
C_{m}^{n}}\left[ L^{\dag }Q_{k-l}^{(k)}
,Q_{m-k+l}^{(n-k)} \right] -\left( n+1\right) L^{\dag }Q_{m+1}^{(n+1)},  \label{eq:Q}
\end{eqnarray}%
\end{widetext}
where $n^{\prime }=\max \left\{ 1,n\right\} ,$ $l_{a}=\max \left\{
0,k-m\right\} ,$ $l_{b}=\min \left\{ k,n-m\right\} ,$ $Q_{m}^{(-1)}
 =Q_{-1}^{(n)} =0$, and $C_{l}^{k}$ is the binomial coefficient.
%
Equations~(\ref{eq:nonlinearSSE}), (\ref{eq:shiftednoisediff}), and (\ref{eq:Q}%
) for $n+m\le \NQmax$ then constitute a set of coupled SODE's from which
$\tilde{\psi}_{t}(\tilde{z}^{\ast })$ can be obtained. To make the results more apparent, we also explicitly show in Appendix~\ref{sec:AppendB} some examples of the evolution equations for the lower order terms $Q_{m}^{(n)}\left(t,\tilde{z}^{\ast}\right)$.

\section{Results on a spin-boson model}

Now we apply our method to a spin-boson model with $H_{\mathrm{sys}}=\frac{\omega}{2}\sigma _{z}$ and $L=\sigma _{x}$ \cite{Lacroix2008,ClosBreuer2012,MakelaMottonen2013,Peropadre2013PRL} 
assuming an initial system state of $\sigmazav=1$ with the bath at zero temperature. In the following calculations, all coupling strengths and frequencies are in units of $\omega$. 
We use $\sqrt{\Gamma\gamma/2}$ to characterize the coupling strength between the system and the environment. This is consistent with the single mode case where the bath spectrum function (i.e., the Lorentzian form) for the Ornstein-Uhlenbeck noise is reduced to $J(\omega)=\Gamma\gamma\delta(\omega)/2$ with $\Gamma\gamma/2$ being the square of the usual single-mode coupling constant. We take $\Gamma\gamma=0.2$ in order to consider the strong coupling regime \cite{FornMooij2010PRL}
(i.e., $\sqrt{\Gamma\gamma/2}\sim0.32\in [0.1, 1]$ in units of $\omega$) and $\gamma=0.2$, $0.4$, and $0.8$ for the bath memory time $1/\gamma$ to show Markovian and non-Markovian behaviors.
Each statistical mean involves an ensemble of $N_z=8000$ of complex colored Gaussian noise $z_t$ obeying the correlation function in Eq.~(\ref{eq:alpha}).
For each realization $z_t$, we obtain one quantum trajectory $\tilde \psi_t(\tilde z_t^\ast)$ by numerically solving the SODE's up to $\NQmax=100$ terms.
The reduced density matrix of the system is recovered by a statistical mean:
$\rho_t=\av{ |\tilde{\psi}_t\left(\tilde z^{\ast }\right)\rangle \langle \tilde{\psi}_t\left(\tilde z^{\ast }\right)|}$.

\begin{figure}[t]
\centering
  \includegraphics[width=.95\columnwidth]{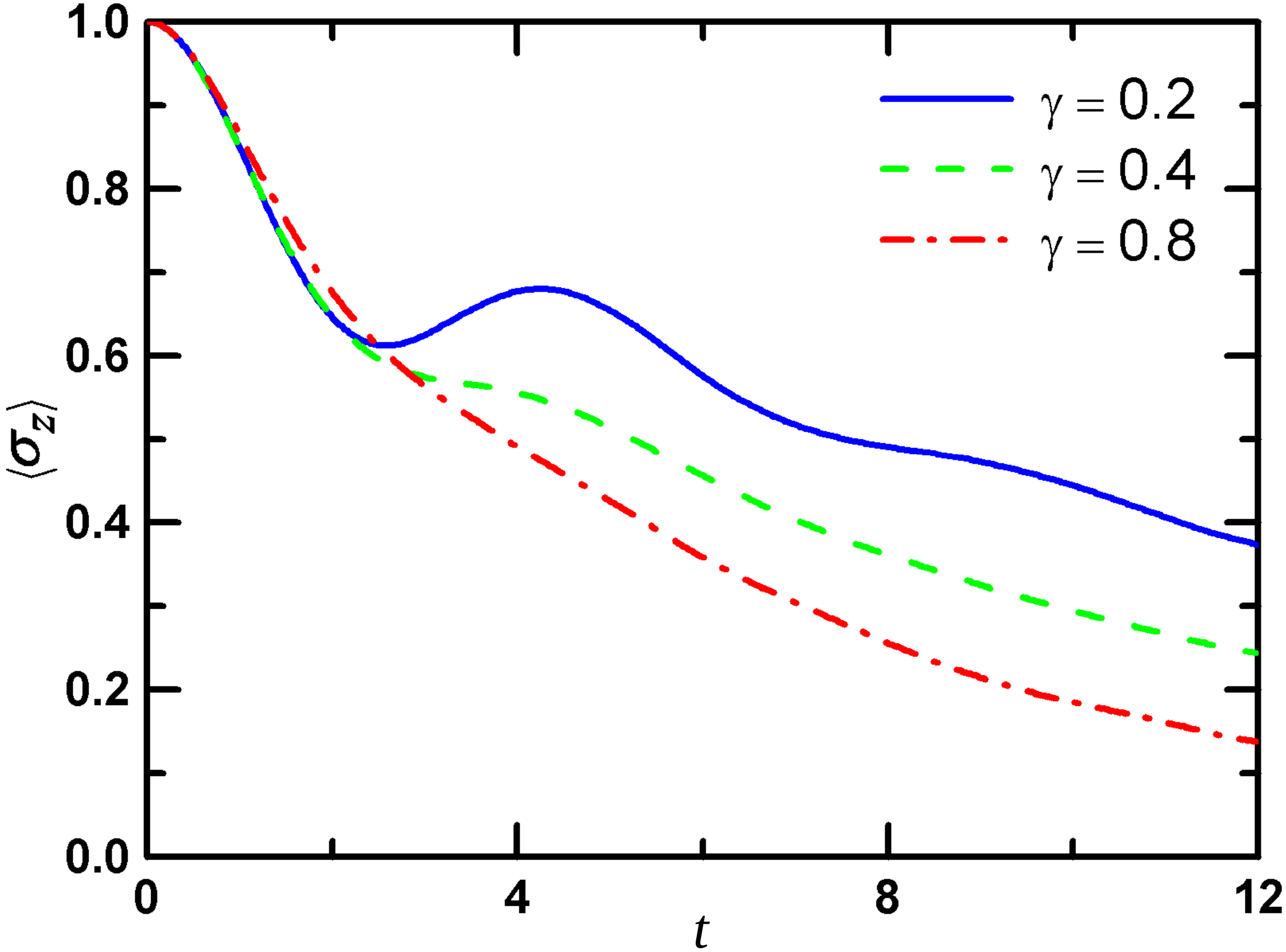}
\caption{(color online) Spin state 
$\langle \sigma_z\rangle$ 
for various memory parameters: $\gamma=0.2$, $0.4$, and $0.8$. Here $\omega=1$, $\Gamma\gamma=0.2$ and $\mathcal{N}=100$ 
.}
\label{fig:nonrwa}
\end{figure}

Figure~\ref{fig:nonrwa} shows the evolutions of $\sigmazav$.
%
For $\gamma=0.2$ corresponding to a relatively long memory time in our study, an oscillatory behavior superimposed with a non-exponentially decay of $\langle \sigma_z\rangle$ is observed, exemplifying strong non-Markovian effects.
The decay behavior becomes more monotonic  as $\gamma$ is increased. At $\gamma=0.8$, it is essentially exponential early on,  demonstrating weak memory effects \cite{Lacroix2008}. In general, exponential decays is ensured when $t\gg 1/\gamma$.
Because the ground state of the total system is no longer a product of the unexcited system state and the vacuum state of the reservoir when including the counter-rotating terms, collapse and revival of the system's state population occur, which indicates that $\langle \sigma_z\rangle$ approaches to zero instead of $-1$ for a long time.
As will be explained below, $\langle \sigma_z\rangle$ reported in Fig.~\ref{fig:nonrwa} admits about $1$\% error.
%

\begin{figure}[t]
\centering
  \includegraphics[width=.95\columnwidth]{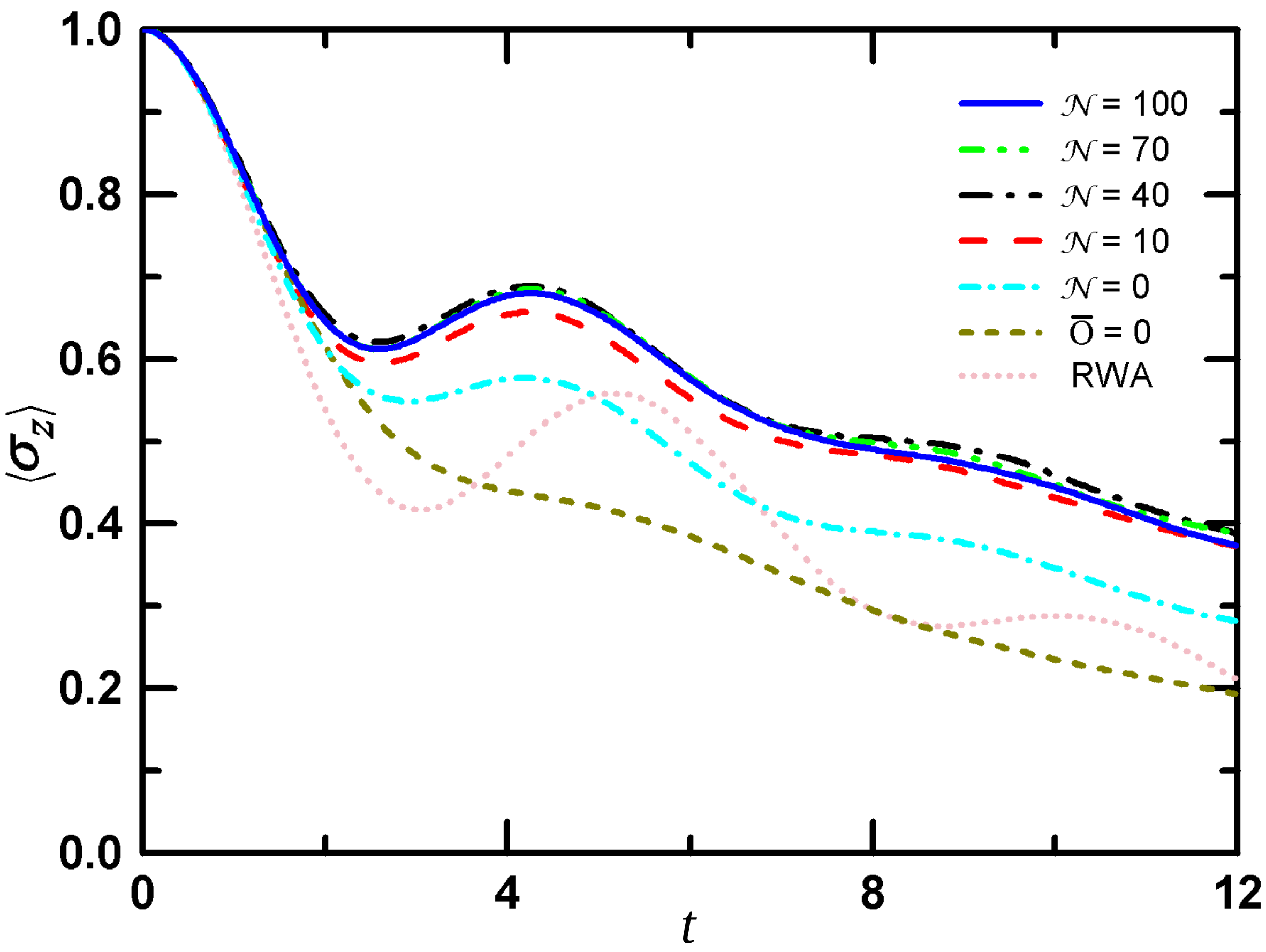}
\caption{(color online) Spin state 
$\langle \sigma_z\rangle$ 
for $\mathcal{N}=100$, $70$, $40$, $10$, and $0$, $\bar{O}=0$, and RWA. Here $\gamma=0.2$ and $\Gamma\gamma=0.2$.}
\label{fig:rwa}
\end{figure}

For comparison, the result for the most interesting case of $\gamma=0.2$ is replotted in Fig.~\ref{fig:rwa} and labelled as $\NQmax=100$. The results for other values of $\NQmax$ are also shown.
We also plot $\sigmazav$ calculated similarly using RWA by taking $L=\sigma _{-}$. RWA is known to be accurate when the system-bath coupling is weak.  At strong coupling considered here, we observe that
the non-Markovian oscillatory behavior of $\langle \sigma_z\rangle$ is successfully reproduced. However, in the RWA, $\sigmazav$ drops considerably faster, 
due to neglecting the counter-rotating terms.  
%

\section{Algorithms and accuracy}

%
\begin{figure}[t]
\centering
\begin{minipage}[c]{0.49\textwidth}
  \centering
  \includegraphics[width=0.95\columnwidth]{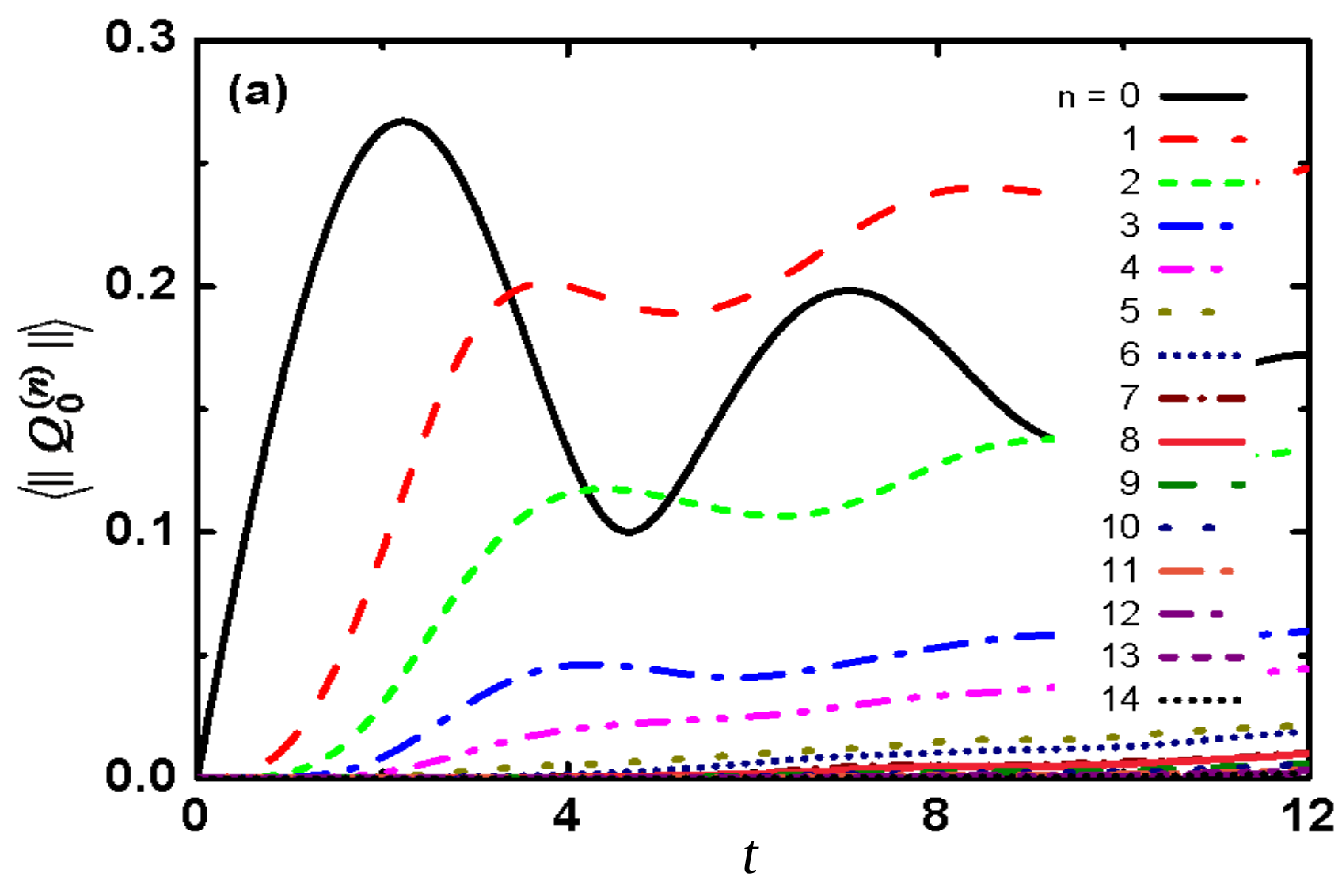}
\end{minipage}\newline
\begin{minipage}[c]{0.49\textwidth}
  \centering
  \includegraphics[width=0.95\columnwidth]{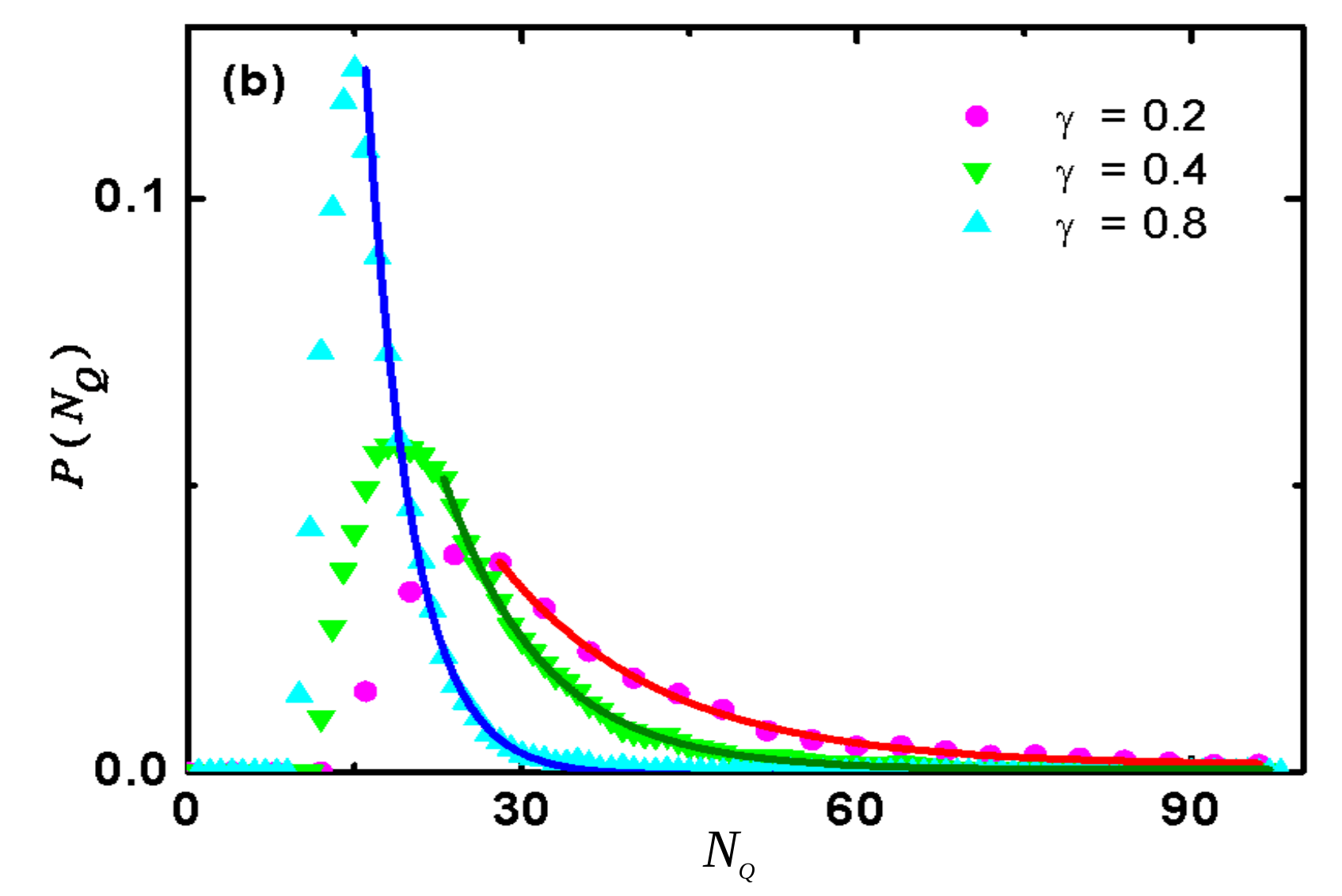}
\end{minipage}
\caption{(color online) (a) 
Average $\langle \parallel Q_0^{\left(n\right)}\parallel\rangle$ of the trace norm of perturbative terms $Q_0^{\left(n\right)}$ 
from the simulation in Fig.~\ref{fig:nonrwa} for $\gamma=0.2$. (b) The probability density $P\left(N_Q\right)$ of $N_Q$ at $t=12$ from the simulations in Fig.~\ref{fig:nonrwa} for $\gamma=0.2$, $0.4$, and $0.8$. Solid lines show fits to exponential distributions for the tails of the distributions.}
\label{fig:NQhistogram}
\end{figure}
Figure~\ref{fig:NQhistogram}(a) plots the 
average $\av{\parallel\Qn \parallel} = \av{  {\rm Tr} \sqrt{\Qn^{\dagger}\Qn} }$ of the trace norm of each perturbative term in Eq.~(\ref{eq:barO}) for $\gamma=0.2$. Initial oscillatory behaviors are observed in the four lowest orders and should be responsible for the similar oscillations in $\sigmazav$.
Moreover, note that the low-order terms rise from 0 earlier than the high-order ones. This verifies that the support of $\Qnm$ expands gradually from low orders as explained above.  For $n\agt 10$, $\av{\parallel\Qn \parallel}$ is already close to 0,  implying good convergence of the functional expansion.
We also observe that $\av{\parallel\Qn\parallel}$ tends to become constant at large $t$. Such saturation is indeed clearly observed for $\gamma=0.4$ and 0.8 at $t\gg 1/\gamma$,  and the saturated value decreases exponentially to 0 with $n$. For any given realization, $\parallel\Qn\parallel$ however persists to fluctuate and arrives only at a dynamic steady state.


To solve the SODE's efficiently, 
we hence put $N_Q=1$ initially and increase it adaptively during the time integration with Euler's method and a time step $\Delta t=0.02$. Only $\Qnm$'s for $n+m\le N_Q$ are calculated and the rest are approximated by zeros. $\Qnm$'s for $n+m=N_Q$ are monitored at every time step. If the magnitude of any of their matrix elements goes beyond a threshold $\epsilon_{\rm thres}=10^{-8}$, $N_Q$ is incremented unless it has reached $\NQmax$,  and the last Euler's step is recalculated.

The number of calculated terms $N_Q$ hence indicates the number of non-zero terms in the functional expansion. It depends on both $t$ and $z_t^{\ast}$,  and hence admits ensemble fluctuations.
Figure~\ref{fig:NQhistogram}(b) plots the probability  density $P(N_Q)$ of its final value at $t=12$ for $N_Q<\NQmax$ from the simulations in Fig.~\ref{fig:nonrwa}.
Interestingly, we observe that the distribution is not narrow. The tails fit very well to exponential forms.
The average $\av{N_Q}$ increases with the memory time $1/\gamma$. Moreover, it also increases with the coupling constant $\Gamma\gamma$ (results not shown).


To study the importance of individual terms, we have also simulated with $\bar{O}=0$ (i.e. $\NQmax<0$), $\NQmax=0$ and $\NQmax=10$. The results are shown in Fig.~\ref{fig:rwa}. We observe as expected that the decay of $\sigmazav$ becomes faster and more monotonic when fewer terms are included. Incidentally, the results for $\bar{O}=0$ and $\NQmax=0$ for $\gamma=0.2$ in Fig.~\ref{fig:rwa} resemble respectively the accurate results for $\gamma=0.8$ and 0.4 in Fig.~\ref{fig:nonrwa}. This suggests that neglecting the non-Markovian terms effectively decreases the bath correlation time.  For $\NQmax=10$, $\sigmazav$ has nearly converged to the accurate result at $\NQmax=100$.

Moreover, for $\gamma=0.2$ and $\NQmax\agt 20$, the solution of the SODE's unexpectedly becomes non-trivial. Once $N_Q$ is constrained at $\NQmax$ and the magnitude of a matrix element of $\Qnm$ with $n+m=\NQmax$ exceeds a tolerance $\epsilon_{\rm tol}=10^{-4}$,
the SODE's eventually become unstable with $Q^{\left(n\right)}_m$ at large $n$ and $m$,  diverging smoothly but rapidly with $t$ even at much reduced $\Delta t$.
The concerned noise realization $z_t^{\ast}$ is hence rejected and excluded from all ensemble averages. Allowing rejection, we have also performed simulations at $\NQmax=40$ and 70 and the results are plotted in Fig.~\ref{fig:rwa}. The rejection rates for $\NQmax=40, 70$, and $100$ are $R=11$\%, $6.6$\% and $5.4$\%,  respectively. Due to the exponential distribution of $N_Q$, $R$ is expected to decrease exponentially against $\NQmax$. We find that the rejected noise realizations $z_t^{\ast}$ in general are those with large magnitudes. The rejection induces errors associated with an ensemble bias which decreases with $\NQmax$. Again, the result for $\NQmax=70$ has nearly converged to our most accurate result at $\NQmax=100$. For $\gamma=0.4$ and 0.8 as shown in Fig.~\ref{fig:nonrwa}, $\av{N_Q}$ is much smaller and thus noise rejection events become rare.

Since the SODE's are exact,  the errors occurring in our algorithm can be fully analyzed.
The r.m.s. error of $\sigmazav$ can be approximated by
$\sqrt{ \mathcal E^2_{Nz} + \mathcal E^2_{\Delta t} + \mathcal
E^2_{\NQmax} }$. Here, $\mathcal E_{Nz} \sim 1/\sqrt{N_z}$ denotes
the ensemble sampling error. For all calculations reported in Fig. 1, we find $\mathcal E_{Nz} \simeq 0.004$ after
averaging over time. The time discretization error $\mathcal
E_{\Delta t}$ is found to be about 0.001 from simulations
with identical noise but different $\Delta t$. Also, $\mathcal
E_{\NQmax}$ is due to including at most $\NQmax=100$ perturbative terms. For $\gamma=0.4$ and
0.8, $\mathcal E_{\NQmax} \simeq 0$ because higher order terms are vanishingly small. For $\gamma=0.2$, we find $\mathcal
E_{\NQmax} \simeq 0.002$ from comparing results at $\NQmax=70$ and 100 with
identical noise. Finally, $\sigmazav$ admits about 1\% error in all three cases.
The simulations for $\gamma=0.2$, 0.4 and 0.8 take about 36, 10, and 2 days respectively to execute on a Intel core-i7 CPU core. Indeed, QSD approaches are fully parallellizable. The accuracy for $\gamma=0.4$ and 0.8 can be further improved substantially by increasing $N_z$. More challenging is the $\gamma=0.2$ case since one must also reduce $\mathcal E_{\NQmax}$ by increasing $\NQmax$. This leads to much more intensive computations. Note that the program run-time is of the order $N_z\NQmax^4 /\Delta t$. Minimizing $\mathcal E_{\NQmax}$ and $\mathcal E_{Nz}$ simultaneously to produce accurate results will be critical and challenging when pushing to even stronger couplings.


\section{Conclusion}

In conclusion, we have developed a high-order non-Markovian QSD approach for open quantum systems based on a set of coupled SODE's which can be
efficiently implemented in numerical simulations.  As an important example, our method is applied to a spin-boson model with a Lorentzian bath spectrum at zero temperature in the strong coupling regime. Note that a generalization to the finite temperature
case is straightforward \cite{Yu2004}. In particular, for this spin-boson model, the finite-temperature non-Markovian
QSD equation actually takes the exact same form as the zero-temperature one.  An extension to general interaction spectra may also be possible by including coupled equations for a full set of new operators analogous to $Q_m^{(n)}$ in Eq.~(\ref{eq:Q_n_m}) each with a particular subset of $\alpha$'s replaced by their derivatives. Our numerical simulations of the spin-boson model have 
shed a new light 
on the spin dynamics without the RWA. It is shown that even though the RWA may successfully reproduce non-Markovian spin-state transient oscillations, it cannot accurately capture the bath memory effects. We emphasize that our proposed approach is efficient and readily applicable to numerically solving the non-Markovian quantum dynamics for open quantum systems with strong coupling and structured bosonic medium. Possible further applications include, for example, multilevel quantum systems in a strong coupling regime~\cite{JingYouYu2012PRA}, photonic band-gap materials~\cite{bandgap,VegaGaspard2005PRA} and also chemical and biological systems~\cite{RodenEisfeldStrunz2009PRL}.

{\it Note added:} Close to the completion of this work, we become aware of a
different kind of numerically exact hierarchical equations by W. Strunz and coworkers \cite{SussEisfeldStrunz14}.

This work is supported by the National Natural Science Foundation of China Grant No.~91121015, the National Basic Research Program of China Grant No.~2014CB921401, the NSAF Grant No.~U1330201, Hong Kong GRF Grant No.~501213, HK PolyU Grant No.~G-YM41, and the China Postdoctoral Science Foundation Grant No.~2012M520146. TY is grateful to Prof.  J. Q. You for the hospitality during his visit to the CSRC, Beijing.

\appendix


\section{Derivation of Eq.~(\ref{eq:Q}) \label{sec:AppendA}}
In this section we provide basic ideas and key derivations to support our central analytical result given by Eq.~(\ref{eq:Q}). Our motivation for a
systematic and efficient approach to solving the non-Markovian quantum state diffusion (QSD) equation is to solve the formidable challenge in the numerical evaluation of the multi-dimensional integrals in the functional expansion of the $O$-operator~\cite{Yu1999PRA}
\begin{eqnarray}
O\left( t,s,\tilde{z}^{\ast}\right) &=&O^{\left( 0\right) }\left( t,s\right)
+\int_{0}^{t}O^{\left( 1\right) }\left( t,s,\upsilon _{1}\right) \tilde{z}^{\ast}_{\upsilon
_{1}}d\upsilon _{1} \notag \\
&&+\int_{0}^{t}\int_{0}^{t}O^{\left( 2\right) }\left(
t,s,\upsilon _{1},\upsilon _{2}\right) \tilde{z}^{\ast}_{\upsilon _{1}}\tilde{z}^{\ast}_{\upsilon
_{2}}d\upsilon _{1}d\upsilon _{2}+\cdots  \notag \\
&&+\int_{0}^{t}\cdots \int_{0}^{t}O^{\left( n\right) }\left( t,s,\upsilon
_{1},\cdots ,\upsilon _{n}\right) \notag \\
&&\times \tilde{z}^{\ast}_{\upsilon _{1}}\cdots \tilde{z}^{\ast}_{\upsilon
_{n}}d\upsilon _{1}\cdots d\upsilon _{n}+\cdots   \label{eq_sm:functional}
\end{eqnarray}
or the $\bar{O}$-operator in Eq.~(\ref{eq:functional}) of the main text. In the notation of Eq.
 ~(\ref{eq:barO}) of the main text, it is clear that only the value of $Q_0^{\left(n\right)}$ contributes to $\bar{O}$ directly. However, in order to have a closed set of equations, we need to introduce more general operators $Q_m^{\left(n\right)}$'s with $n\geq m\geq 0$ which are defined in Eq.
~(\ref{eq:Q_n_m}) 
with $\bar{O}^{\left(n\right)}=\int_0^{t}\alpha\left(t-s\right)O^{\left(n\right)}ds$.
%
To derive readily solvable evolution equations, we differentiate $Q_m^{\left(n\right)}$ w.r.t. time. The calculation is in general straightforward, except that one of the terms contains in the integrand a nontrivial factor
$\int_{0}^{t} \alpha \left( t-s\right) \dot{O}^{\left( n\right) }\left( t,s,\upsilon
_{1},\cdots ,\upsilon _{n}\right) ds$. Using an expression of $\dot{O}^{\left( n\right) }$ from Ref.~\cite{Yu1999PRA}, we obtain
\begin{widetext}
\begin{eqnarray}
\int_{0}^{t} \alpha \left( t-s\right) \dot{O}^{\left( n\right) }\left( t,s,\upsilon
_{1},\cdots ,\upsilon _{n}\right) ds &=&-i\left[H_{S},\bar{O}%
^{\left( n\right) }\left( t,\upsilon _{1},\cdots ,\upsilon _{n}\right) %
\right]  \notag \\
&&-\frac{1}{n!}\sum_{P_{n}\in S_{n}}\sum_{k=0}^{n}\left[ L^{\dag }\bar{O}%
^{\left( k\right) }\left( t,\upsilon _{P_{n}\left( 1\right) },\cdots
,\upsilon _{P_{n}\left( k\right) }\right) ,\bar{O}^{\left( n-k\right)
}\left( t,\upsilon _{P_{n}\left( k+1\right) },\cdots ,\upsilon _{P_{n}\left(
n\right) }\right) \right]  \notag \\
&&-\left( n+1\right) L^{\dag }\int_{0}^{t}\alpha \left( t-\upsilon
_{n+1}\right) \bar{O}^{\left( n+1\right) }\left( t,\upsilon _{1},\cdots
,\upsilon _{n},\upsilon _{n+1}\right) d\upsilon _{n+1}.
\label{eq-sm:Obarequation}
\end{eqnarray}%

In the following we consider cases $n=0$ and $n>0$ separately for the Ornstein-Uhlenbeck noise in Eq.~(\ref{eq:alpha}). For $n=0$, Eq.~(\ref{eq:Q_n_m}) reduces to
\begin{equation}
Q_{0}^{\left( 0\right) }\left( t,\tilde{z}^{\ast}\right) =\bar{O}^{\left(0\right)}\left(t\right) .  \label{eq:Q00}
\end{equation}%
Differentiating Eq.~(\ref{eq:Q00}) gives
\begin{equation}
\dot{Q}_{0}^{\left( 0\right) }\left( t,\tilde{z}^{\ast}\right) =\alpha \left( 0\right)
L-\gamma Q_{0}^{\left( 0\right) }\left( t,\tilde{z}^{\ast}\right) -i\left[
H_{S},Q_{0}^{\left( 0\right) }\left( t,\tilde{z}^{\ast}\right) \right] -\left[ L^{\dag
}Q_{0}^{(0)}\left( t,\tilde{z}^{\ast}\right) ,Q_{0}^{(0)}\left( t,\tilde{z}^{\ast}\right) \right] -L^{\dag
}Q_{0}^{(1)}\left( t,\tilde{z}^{\ast}\right) ,  \label{eq_sm:0thOrder}
\end{equation}
where we have used  Eq.~(\ref{eq-sm:Obarequation}) with $n=0$ and the initial condition  $O^{\left(0\right)}\left(t,t\right)=L$ \cite{Yu1999PRA}.

For $n>0$, we differentiate Eq.~(\ref{eq:Q_n_m}), and obtain
\begin{eqnarray}
\dot{Q}_{m}^{(n)}\left( t,\tilde{z}^{\ast}\right) &=&\partial _{t}Q_{m}^{(n)}\left(
t,\tilde{z}^{\ast}\right)  \notag \\
&=&\left( \partial _{t_{1}}+\partial _{t_{2}}+\partial _{t_{3}}+\partial
_{t_{4}}+\partial _{t_{5}}\right)
\int_{0}^{t_{1}}ds\int_{0}^{t_{2}}d\upsilon _{1}\cdots
\int_{0}^{t_{2}}d\upsilon _{m}\int_{0}^{t_{3}}d\upsilon _{m+1}\cdots
\int_{0}^{t_{3}}d\upsilon _{n}\alpha \left( t_{4}-s\right)  \notag \\
&&\alpha \left( t_{4}-\upsilon _{1}\right) \cdots \alpha \left(
t_{4}-\upsilon _{m}\right) \tilde{z}^{\ast}_{\upsilon _{m+1}}\cdots \tilde{z}^{\ast}_{\upsilon
_{n}}O^{\left( n\right) }\left( t_{5},s,\upsilon _{1},\cdots ,\upsilon
_{n}\right) |_{t_{1}=t_{2}=t_{3}=t_{4}=t_{5}=t}  \notag \\
&=&\frac{m}{n}\alpha \left( 0\right) %
\left[ L,Q_{m-1}^{\left( n-1\right) }\left( t,\tilde{z}^{\ast}\right) \right] +\frac{n-m}{n}%
\tilde{z}^{\ast}_{t}\left[ L,Q_{m}^{\left( n-1\right) }\left( t,\tilde{z}^{\ast}\right) \right] -\left(
m+1\right) \gamma Q_{m}^{\left( n\right) }\left( t,\tilde{z}^{\ast}\right)-i\left[ H_{\mathrm{s}},Q_{m}^{\left( n\right) }\left( t,\tilde{z}^{\ast}\right) %
\right] \notag \\
&&-\sum_{k=0}^{n}\sum_{l=l_{a}}^{l_{b}}\frac{C_{l}^{k}C_{n-m-l}^{n-k}}{%
C_{m}^{n}}\left[ L^{\dag }Q_{k-l}^{(k)}\left( t,\tilde{z}^{\ast}\right)
,Q_{m-k+l}^{(n-k)}\left( t,\tilde{z}^{\ast}\right) \right]-\left( n+1\right) L^{\dag }Q_{m+1}^{\left( n+1\right) }\left( t,\tilde{z}^{\ast}\right) .
\label{eq_sm:non0theOrder}
\end{eqnarray}%
Here we have used Eq.~(\ref{eq-sm:Obarequation}), the symmetric property of $\bar{O}$ with regard to time variables $\upsilon_i$, i.e.
\begin{equation}
\bar{O}^{\left( n\right) }\left( t,\upsilon _{1},\cdots ,\upsilon _{i},\cdots
,\upsilon _{j},\cdots ,\upsilon _{n}\right) =\bar{O}^{\left( n\right) }\left(
t,\upsilon _{1},\cdots ,\upsilon _{j},\cdots ,\upsilon _{i},\cdots
,\upsilon _{n}\right) ,  \label{eq_sm:symmetry}
\end{equation}
and the conditions \cite{Yu1999PRA}
\begin{equation}
O^{\left( n\right) }\left( t,t,\upsilon _{1},\cdots ,\upsilon _{n}\right) =0%
\text{ ~~~~~ }\left( \text{for }n\geqslant 1\right) ,  \label{eq_sm:conditionII}
\end{equation}%
\begin{equation}
O^{\left( n\right) }\left( t,s,t,\upsilon _{2},\cdots ,\upsilon _{n}\right) =%
\frac{1}{n}\left[ L,O^{\left( n-1\right) }\left( t,s,\upsilon _{2},\cdots
,\upsilon _{n}\right) \right] \text{ ~~~~~ }\left( \text{for }n\geqslant
1\right) .  \label{eq_sm:conditionIII}
\end{equation}
Equations~(\ref{eq_sm:0thOrder}) and~(\ref{eq_sm:non0theOrder}) can be combined
and rewritten as Eq.~(9) of the main text.

\section{Examples of $\dot{Q}_m^{\left(n\right)}$ at low orders \label{sec:AppendB}}
In this section, we give some examples of
the evolution equations for the lower order terms ${Q}_{m}^{\left(
n\right)}\left( t,\tilde{z}^{\ast}\right)$ in order to make our results more apparent.
%

$(1)$ When $n=1, m=0$,
\begin{eqnarray}
\dot{Q}_{0}^{(1)}\left( t,\tilde{z}^{\ast}\right) &=&\tilde{z}^{\ast}_{t}\left[ L,Q_{0}^{\left( 0\right)
}\left( t,\tilde{z}^{\ast}\right) \right] -\gamma Q_{0}^{\left( 1\right) }\left( t,\tilde{z}^{\ast}\right)
-i\left[H_{\mathrm{s}},Q_{0}^{\left( 1\right) }\left( t,\tilde{z}^{\ast}\right) %
\right]  \notag \\
&&-\sum_{k=0}^{1}\sum_{l=\max \left\{ 0,k\right\} }^{\min \left\{
k,1\right\} }\frac{C_{l}^{k}C_{1-l}^{1-k}}{C_{0}^{1}}\left[ L^{\dag
}Q_{k-l}^{(k)}\left( t,\tilde{z}^{\ast}\right) ,Q_{-k+l}^{(1-k)}\left( t,\tilde{z}^{\ast}\right) \right]
-2L^{\dag }Q_{1}^{\left( 2\right) }\left( t,\tilde{z}^{\ast}\right)  \notag \\
&=&\tilde{z}^{\ast}_{t}\left[ L,Q_{0}^{\left( 0\right) }\left( t,\tilde{z}^{\ast}\right) \right] -\gamma
Q_{0}^{\left( 1\right) }\left( t,\tilde{z}^{\ast}\right) -i\left[
H_{\mathrm{s}},Q_{0}^{\left( 1\right) }\left( t,\tilde{z}^{\ast}\right) \right] -\left[ L^{\dag
}Q_{0}^{(0)}\left( t,\tilde{z}^{\ast}\right) ,Q_{0}^{(1)}\left( t,\tilde{z}^{\ast}\right) \right]  \notag
\\
&&-\left[ L^{\dag }Q_{0}^{(1)}\left( t,\tilde{z}^{\ast}\right) ,Q_{0}^{(0)}\left(
t,\tilde{z}^{\ast}\right) \right] -2L^{\dag }Q_{1}^{\left( 2\right) }\left( t,\tilde{z}^{\ast}\right) .
\end{eqnarray}

$(2)$ When $n=m=1$,
\begin{eqnarray}
\dot{Q}_{1}^{(1)}\left( t,\tilde{z}^{\ast}\right) &=&\alpha \left( 0\right) \left[
L,Q_{0}^{\left( 0\right) }\left( t\right) \right] -2\gamma Q_{1}^{\left(
1\right) }\left( t,\tilde{z}^{\ast}\right) -i\left[H_{\mathrm{s}},Q_{1}^{\left(
1\right) }\left( t,\tilde{z}^{\ast}\right) \right]  \notag \\
&&-\sum_{k=0}^{1}\sum_{l=\max \left\{ 0,k-1\right\} }^{\min \left\{
k,0\right\} }\frac{C_{l}^{k}C_{0-l}^{1-k}}{C_{1}^{1}}\left[ L^{\dag
}Q_{k-l}^{(k)}\left( t,\tilde{z}^{\ast}\right) ,Q_{1-k+l}^{(1-k)}\left( t,\tilde{z}^{\ast}\right) \right]
-2L^{\dag }Q_{2}^{\left( 2\right) }\left( t,\tilde{z}^{\ast}\right)  \notag \\
&=&\alpha \left( 0\right) \left[ L,Q_{0}^{\left( 0\right) }\left( t\right) %
\right] -2\gamma Q_{1}^{\left( 1\right) }\left( t,\tilde{z}^{\ast}\right) -i\left[
H_{\mathrm{s}},Q_{1}^{\left( 1\right) }\left( t,\tilde{z}^{\ast}\right) \right] -\left[
L^{\dag }Q_{0}^{(0)}\left( t,\tilde{z}^{\ast}\right) ,Q_{1}^{(1)}\left( t,\tilde{z}^{\ast}\right) \right]
\notag \\
&&-\left[ L^{\dag }Q_{1}^{(1)}\left( t,\tilde{z}^{\ast}\right) ,Q_{0}^{(0)}\left(
t,\tilde{z}^{\ast}\right) \right] -2L^{\dag }Q_{2}^{\left( 2\right) }\left( t,\tilde{z}^{\ast}\right) .
\end{eqnarray}

$(3)$ When $n=2, m=1,$
\begin{equation*}
\dot{Q}_{1}^{(2)}\left( t,\tilde{z}^{\ast}\right)=\frac{\alpha\left( 0\right)}{2} \left[
L,Q_{0}^{\left( 1\right) }\left( t,\tilde{z}^{\ast}\right) \right]+\frac{\tilde{z}^{\ast}_{t}}{2}\left[
L,Q_{1}^{\left( 1\right) }\left( t,\tilde{z}^{\ast}\right) \right] -2\gamma Q_{1}^{\left(
2\right) }\left( t,\tilde{z}^{\ast}\right) -i\left[H_{\mathrm{s}},Q_{1}^{\left(
2\right) }\left( t,\tilde{z}^{\ast}\right) \right] -\left\{ \sum_{k=0}^{2} \mathcal{E}%
_k\right\} -3L^{\dag }Q_{2}^{\left( 3\right) }\left( t,\tilde{z}^{\ast}\right) ,
\end{equation*}
where
\begin{eqnarray}
\left\{ \sum_{k=0}^{2} \mathcal{E}_k\right\}&=&\frac{1}{2!}%
\int_{0}^{t}d\upsilon _{1}\int_{0}^{t}d\upsilon _{2}\alpha \left( t-\upsilon
_{1}\right) \tilde{z}^{\ast}_{\upsilon _{2}}\sum_{P_{2}\in S_{2}}\sum_{k=0}^{2}\left[
L^{\dag }\bar{O}^{\left( k\right) }\left( t,\upsilon _{P_{2}\left( 1\right)
},...,\upsilon _{P_{2}\left( k\right) }\right) ,\bar{O}^{\left( 2-k\right)
}\left( t,\upsilon _{P_{2}\left( k+1\right) },...,\upsilon _{P_{2}\left(
2\right) }\right) \right]  \notag \\
&=&\frac{1}{2!}\int_{0}^{t}d\upsilon _{1}\int_{0}^{t}d\upsilon _{2}\alpha
\left( t-\upsilon _{1}\right) \tilde{z}^{\ast}_{\upsilon _{2}}\Bigg\{2\left[ L^{\dag }\bar{O%
}^{\left( 0\right) }\left( t\right) ,\bar{O}^{\left( 2\right) }\left(
t,\upsilon _{1},\upsilon _{2}\right) \newline
\right] +\left[ L^{\dag }\bar{O}^{\left( 1\right) }\left( t,\upsilon
_{1}\right) ,\bar{O}^{\left( 1\right) }\left( t,\upsilon _{2}\right) \right]
\notag \\
&&+\left[ L^{\dag }\bar{O}^{\left( 1\right) }\left( t,\upsilon _{2}\right) ,%
\bar{O}^{\left( 1\right) }\left( t,\upsilon _{1}\right) \right] +2\left[
L^{\dag }\bar{O}^{\left( 2\right) }\left( t,\upsilon _{1},\upsilon
_{2}\right) ,\bar{O}^{\left( 0\right) }\left( t\right) \right] \Bigg\}
\notag \\
&=&\sum_{k=0}^{2}\sum_{l=\max \left\{ 0,k-1\right\} }^{\min \left\{
k,1\right\} }\frac{C_{l}^{k}C_{1-l}^{2-k}}{C_{1}^{2}}\left[ L^{\dag
}Q_{k-l}^{(k)}\left( t,\tilde{z}^{\ast}\right) ,Q_{1-k+l}^{(2-k)}\left( t,\tilde{z}^{\ast}\right) \right].
\end{eqnarray}%
Therefore,
\begin{eqnarray}
\dot{Q}_{1}^{(2)}\left( t,\tilde{z}^{\ast}\right)&=&\frac{\alpha\left( 0\right)}{2}\left[
L,Q_{0}^{\left( 1\right) }\left( t,\tilde{z}^{\ast}\right) \right]+\frac{\tilde{z}^{\ast}_{t}}{2}\left[
L,Q_{1}^{\left( 1\right) }\left( t,\tilde{z}^{\ast}\right) \right] -2\gamma Q_{1}^{\left(
2\right) }\left( t,\tilde{z}^{\ast}\right) -i\left[H_{\mathrm{s}},Q_{1}^{\left(
2\right) }\left( t,\tilde{z}^{\ast}\right) \right] -3L^{\dag }Q_{2}^{\left( 3\right)
}\left( t,\tilde{z}^{\ast}\right)  \notag \\
&&-\frac{1}{2}\Bigg\{2\left[ L^{\dag }Q_{0}^{\left( 0\right)
}\left( t,\tilde{z}^{\ast}\right) ,Q_{1}^{\left( 2\right) }\left( t,\tilde{z}^{\ast}\right) %
\right] +\left[ L^{\dag }Q_{1}^{\left( 1\right) }\left(
t,\tilde{z}^{\ast}\right) ,Q_{0}^{\left( 1\right) }\left( t,\tilde{z}^{\ast}\right) \right] +%
\left[ L^{\dag }Q_{0}^{\left( 1\right) }\left( t,\tilde{z}^{\ast}\right)
,Q_{1}^{\left( 1\right) }\left( t,\tilde{z}^{\ast}\right) \right]  \notag \\
&&+2\left[L^{\dag }Q_{ 1 }^{\left( 2\right) }\left( t,\tilde{z}^{\ast}\right)
,Q_{0 }^{\left( 0\right) }\left( t,\tilde{z}^{\ast}\right) \right] \Bigg\}.
\end{eqnarray}%

$(4)$ When $n=m=2$,
\begin{eqnarray}
\dot{Q}_{2}^{(2)}\left( t,\tilde{z}^{\ast}\right) &=&\alpha \left( 0\right) \left[
L,Q_{1}^{\left( 1\right) }\left( t,\tilde{z}^{\ast}\right) \right] -3\gamma Q_{2}^{\left(
2\right) }\left( t,\tilde{z}^{\ast}\right) -i\left[H_{\mathrm{s}},Q_{2}^{\left(
2\right) }\left( t,\tilde{z}^{\ast}\right) \right]  \notag \\
&&-\sum_{k=0}^{2}\sum_{l=\max \left\{ 0,k-2\right\} }^{\min \left\{
k,0\right\} }\frac{C_{l}^{k}C_{0-l}^{2-k}}{C_{2}^{2}}\left[ L^{\dag
}Q_{k-l}^{(k)}\left( t,\tilde{z}^{\ast}\right) ,Q_{2-k+l}^{(2-k)}\left( t,\tilde{z}^{\ast}\right) \right]
-3L^{\dag }Q_{3}^{\left( 3\right) }\left( t,\tilde{z}^{\ast}\right)  \notag \\
&=&\alpha \left( 0\right)\left[ L,Q_{1}^{\left( 1\right) }\left( t,\tilde{z}^{\ast}\right) %
\right] -3\gamma Q_{2}^{\left( 2\right) }\left( t,\tilde{z}^{\ast}\right) -i\left[H_{\mathrm{s}},Q_{2}^{\left( 2\right) }\left( t,\tilde{z}^{\ast}\right) \right] -\left[
L^{\dag }Q_{0}^{(0)}\left( t,\tilde{z}^{\ast}\right) ,Q_{2}^{(2)}\left( t,\tilde{z}^{\ast}\right) \right]
\notag \\
&&-\left[ L^{\dag }Q_{1}^{(1)}\left( t,\tilde{z}^{\ast}\right) ,Q_{1}^{(1)}\left(
t,\tilde{z}^{\ast}\right) \right] -\left[ L^{\dag }Q_{2}^{(2)}\left( t,\tilde{z}^{\ast}\right)
,Q_{0}^{(0)}\left( t,\tilde{z}^{\ast}\right) \right] -3L^{\dag }Q_{3}^{\left( 3\right)
}\left( t,\tilde{z}^{\ast}\right) .
\end{eqnarray}

$(5)$ When $n=3, m=1$,
\begin{eqnarray}
\dot{Q}_{1}^{(3)}\left( t,\tilde{z}^{\ast}\right) &=&\frac{1}{3}\alpha \left( 0\right) %
\left[ L,Q_{0}^{\left( 2\right) }\left( t,\tilde{z}^{\ast}\right) \right] +\frac{2}{3}z_{t}%
\left[ L,Q_{1}^{\left( 2\right) }\left( t,\tilde{z}^{\ast}\right) \right] -2\gamma
Q_{1}^{\left( 3\right) }\left( t,\tilde{z}^{\ast}\right) -i\left[
H_{\mathrm{s}},Q_{1}^{\left( 3\right) }\left( t,\tilde{z}^{\ast}\right) \right]  \notag \\
&&-\sum_{k=0}^{3}\sum_{l=\max \left\{ 0,k-1\right\} }^{\min \left\{
k,2\right\} }\frac{C_{l}^{k}C_{2-l}^{3-k}}{C_{1}^{3}}\left[ L^{\dag
}Q_{k-l}^{(k)}\left( t,\tilde{z}^{\ast}\right) ,Q_{1-k+l}^{(3-k)}\left( t,\tilde{z}^{\ast}\right) \right]
-4L^{\dag }Q_{2}^{\left( 4\right) }\left( t,\tilde{z}^{\ast}\right)  \notag \\
&=&\frac{1}{3}\alpha \left( 0\right) \left[ L,Q_{0}^{\left( 2\right) }\left(
t,\tilde{z}^{\ast}\right) \right] +\frac{2}{3}z_{t}\left[ L,Q_{1}^{\left( 2\right) }\left(
t,\tilde{z}^{\ast}\right) \right] -2\gamma Q_{1}^{\left( 3\right) }\left( t,\tilde{z}^{\ast}\right) -i\left[
H_{\mathrm{s}},Q_{1}^{\left( 3\right) }\left( t,\tilde{z}^{\ast}\right) \right]
\notag \\
&&-\left[ L^{\dag }Q_{0}^{(0)}\left( t,\tilde{z}^{\ast}\right) ,Q_{1}^{(3)}\left(
t,\tilde{z}^{\ast}\right) \right] -\frac{1}{3}\left[ L^{\dag }Q_{1}^{(1)}\left( t,\tilde{z}^{\ast}\right)
,Q_{0}^{(2)}\left( t,\tilde{z}^{\ast}\right) \right] -\frac{2}{3}\left[ L^{\dag
}Q_{0}^{(1)}\left( t,\tilde{z}^{\ast}\right) ,Q_{1}^{(2)}\left( t,\tilde{z}^{\ast}\right) \right]  \notag
\\
&&-\frac{2}{3}\left[ L^{\dag }Q_{1}^{(2)}\left( t,\tilde{z}^{\ast}\right)
,Q_{0}^{(1)}\left( t,\tilde{z}^{\ast}\right) \right] -\frac{1}{3}\left[ L^{\dag
}Q_{0}^{(2)}\left( t,\tilde{z}^{\ast}\right) ,Q_{1}^{(1)}\left( t,\tilde{z}^{\ast}\right) \right] -\left[
L^{\dag }Q_{1}^{(3)}\left( t,\tilde{z}^{\ast}\right) ,Q_{0}^{(0)}\left( t,\tilde{z}^{\ast}\right) \right]
\notag \\
&&-4L^{\dag }Q_{2}^{\left( 4\right) }\left( t,\tilde{z}^{\ast}\right) .
\end{eqnarray}

$(6)$ When $n=3, m=2$,
\begin{eqnarray}
\dot{Q}_{2}^{(3)}\left( t,\tilde{z}^{\ast}\right) &=&\frac{2}{3}\alpha \left( 0\right) %
\left[ L,Q_{1}^{\left( 2\right) }\left( t,\tilde{z}^{\ast}\right) \right] +\frac{1}{3}\tilde{z}^{\ast}_{t}%
\left[ L,Q_{2}^{\left( 2\right) }\left( t,\tilde{z}^{\ast}\right) \right] -3\gamma
Q_{2}^{\left( 3\right) }\left( t,\tilde{z}^{\ast}\right) -i\left[
H_{\mathrm{s}},Q_{2}^{\left( 3\right) }\left( t,\tilde{z}^{\ast}\right) \right]  \notag \\
&&-\sum_{k=0}^{3}\sum_{l=\max \left\{ 0,k-2\right\} }^{\min \left\{
k,1\right\} }\frac{C_{l}^{k}C_{1-l}^{3-k}}{C_{2}^{3}}\left[ L^{\dag
}Q_{k-l}^{(k)}\left( t,\tilde{z}^{\ast}\right) ,Q_{2-k+l}^{(3-k)}\left( t,\tilde{z}^{\ast}\right) \right]
-4L^{\dag }Q_{3}^{\left( 4\right) }\left( t,\tilde{z}^{\ast}\right)  \notag \\
&=&\frac{2}{3}\alpha \left( 0\right) \left[ L,Q_{1}^{\left( 2\right) }\left(
t,\tilde{z}^{\ast}\right) \right] +\frac{1}{3}\tilde{z}^{\ast}_{t}\left[ L,Q_{2}^{\left( 2\right) }\left(
t,\tilde{z}^{\ast}\right) \right] -3\gamma Q_{2}^{\left( 3\right) }\left( t,\tilde{z}^{\ast}\right) -i\left[
H_{\mathrm{s}},Q_{2}^{\left( 3\right) }\left( t,\tilde{z}^{\ast}\right) \right]
\notag \\
&&-\left[ L^{\dag }Q_{0}^{(0)}\left( t,\tilde{z}^{\ast}\right) ,Q_{2}^{(3)}\left(
t,\tilde{z}^{\ast}\right) \right] -\frac{2}{3}\left[ L^{\dag }Q_{1}^{(1)}\left( t,\tilde{z}^{\ast}\right)
,Q_{1}^{(2)}\left( t,\tilde{z}^{\ast}\right) \right] -\frac{1}{3}\left[ L^{\dag
}Q_{0}^{(1)}\left( t,\tilde{z}^{\ast}\right) ,Q_{2}^{(2)}\left( t,\tilde{z}^{\ast}\right) \right]  \notag
\\
&&-\frac{1}{3}\left[ L^{\dag }Q_{2}^{(2)}\left( t,\tilde{z}^{\ast}\right)
,Q_{0}^{(1)}\left( t,\tilde{z}^{\ast}\right) \right] -\frac{2}{3}\left[ L^{\dag
}Q_{1}^{(2)}\left( t,\tilde{z}^{\ast}\right) ,Q_{1}^{(1)}\left( t,\tilde{z}^{\ast}\right) \right] -\left[
L^{\dag }Q_{2}^{(3)}\left( t,\tilde{z}^{\ast}\right) ,Q_{0}^{(0)}\left( t,\tilde{z}^{\ast}\right) \right]
\notag \\
&&-4L^{\dag }Q_{3}^{\left( 4\right) }\left( t,\tilde{z}^{\ast}\right) .
\end{eqnarray}

$(7)$ When $n=m=3$,
\begin{eqnarray}
\dot{Q}_{3}^{(3)}\left( t,\tilde{z}^{\ast}\right) &=&\alpha \left( 0\right) \left[
L,Q_{2}^{\left( 2\right) }\left( t,\tilde{z}^{\ast}\right) \right] -4\gamma Q_{3}^{\left(
3\right) }\left( t,\tilde{z}^{\ast}\right)  \notag \\
&&-i\left[H_{\mathrm{s}},Q_{3}^{\left( 3\right) }\left( t,\tilde{z}^{\ast}\right) %
\right] -\sum_{k=0}^{3}\sum_{l=\max \left\{ 0,k-3\right\} }^{\min \left\{
k,0\right\} }\frac{C_{l}^{k}C_{0-l}^{3-k}}{C_{3}^{3}}\left[ L^{\dag
}Q_{k-l}^{(k)}\left( t,\tilde{z}^{\ast}\right) ,Q_{3-k+l}^{(3-k)}\left( t,\tilde{z}^{\ast}\right) \right]
-4L^{\dag }Q_{4}^{\left( 4\right) }\left( t,\tilde{z}^{\ast}\right)  \notag \\
&=&\alpha \left( 0\right) \left[ L,Q_{2}^{\left( 2\right) }\left( t,\tilde{z}^{\ast}\right) %
\right] -4\gamma Q_{3}^{\left( 3\right) }\left( t,\tilde{z}^{\ast}\right) -i\left[H_{\mathrm{s}},Q_{3}^{\left( 3\right) }\left( t,\tilde{z}^{\ast}\right) \right] -\left[
L^{\dag }Q_{0}^{(0)}\left( t,\tilde{z}^{\ast}\right) ,Q_{3}^{(3)}\left( t,\tilde{z}^{\ast}\right) \right]
\notag \\
&&-\left[ L^{\dag }Q_{1}^{(1)}\left( t,\tilde{z}^{\ast}\right) ,Q_{2}^{(2)}\left(
t,\tilde{z}^{\ast}\right) \right] -\left[ L^{\dag }Q_{2}^{(2)}\left( t,\tilde{z}^{\ast}\right)
,Q_{1}^{(1)}\left( t,\tilde{z}^{\ast}\right) \right] -\left[ L^{\dag }Q_{3}^{(3)}\left(
t,\tilde{z}^{\ast}\right) ,Q_{0}^{(0)}\left( t,\tilde{z}^{\ast}\right) \right]  \notag \\
&&-4L^{\dag }Q_{4}^{\left( 4\right) }\left( t,\tilde{z}^{\ast}\right) .
\end{eqnarray}
\end{widetext}

\end{document}